\documentclass[preprint,aps,showpacs]{revtex4}

\usepackage{graphicx}

\begin{document}

\title
{ Pair-distribution functions 
 of two-temperature two-mass systems: 
Comparison of MD, HNC, CHNC, QMC and
Kohn-Sham calculations for dense hydrogen
}

\author
{
 M.W.C. Dharma-wardana}
\affiliation{
National Research Council of Canada, Ottawa, Canada, K1A 0R6
}
\email[Email address:\ ]{chandre.dharma-wardana@nrc-cnrc.gc.ca}

\author
{Michael S. Murillo}
\affiliation{Physics Division, MS D410, Los Alamos National Laboratory, Los Alamos, New Mexico, 87545}
\email[Email address:\ ]{murillo@lanl.gov}
%
\date{\today}

\begin{abstract}
Two-temperature, two-mass quasi-equilibrium plasmas may occur in electron-ion plasmas,
nuclear-matter, as well as in electron-hole condensed-matter systems. 
Dense two-temperature hydrogen plasmas straddle the difficult partially-degenerate
 regime
of electron densities and temperatures which are important in astrophysics,  in
 inertial-confinement
fusion research, and other areas of warm dense matter physics. Results from
 Kohn-Sham calculations and QMC are used to benchmark the procedures used in
classical molecular-dynamics simulations, HNC and CHNC methods to
derive electron-electron and electron-proton pair-distribution functions. 
 Then, nonequilibrium
molecular dynamics for two-temperature, two-mass plasmas are used to
 obtain the pair distribution.
  Using these results, the correct HNC and 
CHNC procedures for the evaluation of pair-distribution functions in
 two-temperature two-mass two-component
charged fluids are established. Results for a mass ratio of 1:5, typical of
electron-hole fluids, as well as for compressed hydrogen are presented.
\end{abstract}
%
\maketitle
\section{Introduction}
 The study of hot strongly-coupled dense charged
 fluids is a difficult task, especially near the
  regime of or molecular and atomic species\cite{ceperleyH}, or excitons in electron-hole plasmas.
   The system is better understood
  for fully ionized systems, such as hydrogen, in the form of free electrons
  and protons, and fully-ionized electron-hole condensates.
   In fact, considerable headway 
 has been made using methods based on density-functional theory (DFT), even
 for plasmas with multiple states of ionization.
 DFT methods have been used with molecular-dynamics based
 approaches\cite{kwon,dejarlais,mazevet}, as well as within multi-component
 integral-equation approaches \cite{pdw95,ilciacco}. Equilibrium properties
 of plasmas, as well as their linear transport properties, have been
 successfully studied in these papers, and excellent agreement
 between the molecular-dynamics based DFT and integral-equation
 based DFT has been found \cite{lvm}. 

On the other hand, laser-produced plasmas are initially formed as
two-temperature plasmas, where the electrons have absorbed the laser
energy and have self-equilibrated to some ``electron temperature''
$T_e$, while the ions remain cool, at some temperature $T_i$, with
$T_i<T_e$. The opposite
situation arises in shock-wave generated plasmas, where the ions absorb the
shock energy and $T_i>T_e$. Such two-temperature plasmas also occur in
astrophysical settings, affecting the time of termination
 of synthesis of light-nuclei
to occur at different stages of cooling of the electrons\cite{astrop}, and  
influencing the Coulomb nuclear-tunneling rates\cite{dewitt}.
The possibility of such well-defined 
two-temperature plasmas is largely a result of the extreme mass ratio
$m_i/m_e\ge 1836$ between ions and electrons. Similar, but less well
defined situations can arise
 in electron-hole plasmas, where the masses are of the same order of
 magnitude (e.g, the electron and hole masses in GaAs are 0.067$m_e$ and
 0.34$m_e$ respectively, with an electron-hole mass ratio of $\sim 5$).
 GaAs is a direct bandgap material, and electron-hole plasmas are more
 easily studied in indirect-gap systems like Si where the density-of states
 mass ratio is $\sim 3$. 
  The simulation of such systems at two temperatures, using quantum Monte-Carlo
methods is at present unavailable, even in regimes of densities
and temperatures where bound states (or exciton formation in electron-hole
systems) do not exist. Thus it is natural to look for analytical methods based on
integral-equation techniques which are computationally simple
and physically insightful. However, although $T_e,\, T_i$ define
 the temperatures of each subsystem and the pair-distribution
 functions (PDFs) $g_{ee}$ and $g_{ii}$, the  `temperature' $T_{ei}$
entering into
 the cross-correlations $g_{ei}$,  as well as the the effects
 of electron spin, exchange 
 etc., relevant to two-temperature systems need to be clarified. In this context
 we use $T_{ee}=T_e$, $T_{ii}=T_i$, and $T_{ei}$ to refer to the
 electron-, ion-, and electron-ion temperatures as they enter independently
 into the Ornstein-Zernike (OZ) and hypernetted chain (HNC) equations. In fact,
 some authors\cite{seuf} have proposed to modify
  the well-established
 OZ equations in dealing with
  two-temperature (2T) two-mass (2M) systems.

The objective of this paper is to study such 2T-2M  plasmas
using results from molecular-dynamics (MD)\cite{mc}, HNC\cite{hncref},
classical-map HNC (CHNC)\cite{prl1,prb3d}, quantum Monte Carlo (QMC)\cite{mc}
 and Kohn-Sham (KS)\cite{ilciacco} methods to establish the
proper implementation of quantum  effects and 2T, 2M situations in
 simulation studies.
One of our main interests would be uniform hydrogenic plasmas free of
bound states, in the regime of warm-dense matter. 
\section{Theoretical Methods}
A system of classical particles, e.g., hard-spheres or Lennard-Jones fluids,
 or classical
ions in a uniform neutralizing background, can
be studied completely using the method of molecular dynamics (MD) where
 the classical equations
of motion are integrated numerically, for a sufficiently large number of
 particles contained
in a simulation box. It has been found that the particle distribution functions,
e.g., the pair-distribution function (PDF) $g_{ij}(r)$ (where $i,j$ 
 specify the species, spin etc.),
 obtained from MD simulations for charged classical ions
can also be accurately reproduced via suitable integral-equations which are
computationally very economical and efficient.
 The pair-potentials, quantum corrections etc.,
needed to simulate systems with ions and electrons, or purely electron systems
 (with a uniform neutralizing background) will be discussed in this section.  First, we compare
the usual HNC approximation with the CHNC method.  Then, these results are compared with a 
full quantum (Kohn-Sham) calculation; these results are then used to determine the effective 
diffractive interaction used in the CHNC method.  Finally, we discuss how the CHNC can be extended to 
``classical map molecular dynamics'' (CMMD).
\subsection{HNC and CHNC methods}
The HNC equation and
its straight-forward generalizations, coupled with the  OZ equation
 have lead
to very accurate results for classical charged-particle interactions.
 The exact equations for the PDFs are of the form:
\begin{equation}
g_{ij}(r)=e^{-\beta_{cf} \phi_{ij}(r)
+h_{ij}(r)-c_{ij}(r) + B_{ij}(r)}
\label{hnc}
\end{equation}
Here $\phi_{ij}(r)$ is the pair potential between the
species $i,j$.
If the bridge function $B_{ij}(r)$ is set to zero we have the HNC approximation.
Then, given the temperature $T=1/\beta$, the particle densities $n_i$, and
the pair-potentials $\phi_{ij}(r)$, the pair-correlation function $h_{ij}(r)$ and
the direct correlation function $c_{ij}(r)$ can be self-consistently obtained via the
HNC and OZ equations, which have the form
\begin{equation}
\label{oz1}
 h_{ij}(r)  = c_{ij}(r)+
\sum_ s n_s\int d{\bf r}'h_{is}
(|{\bf r}-{\bf r}'|)c_{sj}({\bf r}').
\end{equation}
This method already fails for strictly attractive potentials. Thus,
in simulations of electron-proton systems, $\phi_{ep}(r)$ has to
be replaced by effective-potentials which attempt to incorporate
quantum diffraction effects \cite{Jones}. 
Even with purely repulsive potentials, classical simulations fail to
incorporate Fermi or Bose statistics which begin to
manifest as the density is increased and
the temperature is lowered. A well established means of
incorporating quantum effects is to derive
integral equations from correlated-determinantal
 wavefunctions, as done in the Feenberg
approach\cite{feenberg,lantto}. The resulting integral equations
 are very daunting, and
in fact, not easy to use. Quantum Monte-Carlo (QMC) itself may be considered
as an adaptation of the Feenberg-functional to generate a statistical measure
 for the stochastic algorithms used in MD. An alternative approach, using
  Feynman paths instead of classical trajectories, provides another class
 of simulation techniques. However, these
 quantum simulation methods become computationally extremely 
heavy. Such methods are
best suited for the establishment of bench-mark results, and for the
 ``calibration'' of
other methods which contain approximation schemes. In fact, the QMC 
techniques have been most useful in providing the ``exchange-correlation''
 potentials $V_{xc}(r)$ needed in the Kohn-Sham
density-functional theory (DFT) equations. While DFT is itself exact,
one has to use results from QMC and such microscopic methods to model
the unknown $V_{xc}(r)$. Given the $V_{xc}$, the inhomogeneous density distribution
around a given particle can be calculated, and the pair-distribution is deduced from
it.

The method followed here is to exploit the well established Kohn-Sham
equations as the reference calculation, and determine the effective
potentials to be used in the classical simulations of 2T and 2M systems.
 To this end we present comparisons of Kohn-Sham calculations of $g_{ep}(r)$
 for
 H-plasmas with available QMC results to mutually validate these methods.
  Another theoretical tool we use is the ``classical-map
HNC'', i.e., CHNC equations which incorporate quantum effects including
Fermion statistics via effective potentials and effective temperatures.
The CHNC has been extensively tested via comparisons
 with QMC results in 2-D and 3-D electron systems, and shown to provide excellent
  agreement, even at the extreme quantum limit of zero temperature\cite{prl1, prl2}.
    CHNC uses
 a ``quantum temperature'' $T_q$, which depends on the Fermion density. If the
 physical temperature of the quantum fluid is $T$, the distribution functions
  are obtained\cite{prl1} from a
classical fluid at the temperature $T_{cf}$ such that:
\begin{equation}
\label{tcf}
T_{cf}=\sqrt{T^2+T_q^2}.
\end{equation}
The temperature $T_q$ is defined to be such that the classical Coulomb fluid has the same
exchange-correlation
 energy as the
quantum fluid [\onlinecite{prl1}]. DFT assures us that only
 the true density
distribution possesses the true exchange-correlation energy.
 Thus, the charge distributions, i.e., the
PDFs obtained from CHNC are found to be in excellent agreement with those
 from Monte-Carlo simulations of 2D and 3D systems.
This agreement is obtained by including the exchange-hole of parallel-spin
 electrons as an
effective potential, called in CHNC the Pauli exclusion potential $P_{ij}(r)$.
 Clearly, this is zero if $i\neq j$. For $i=j=\parallel$ spins, the potential $P_{ij}(r)$ is such that the
 non-interacting PDFs, i.e.,
$g_{ij}^0(r)$ are correctly recovered from the integral equations\cite{lado}.
 Thus, using atomic units where $|e|=\hbar=m_e=1$, the effective
pair potentials $\phi_{ij}(r)$ are of the form:
\begin{eqnarray}
\label{pots}
\phi_{ij}(r)&=&P_{ij}(r)+ V^c_{ij}(r)\\
V^c_{ij}(r)&=& z_iz_j(1-e^{-k_{ij}r})/r
\end{eqnarray}
Here $z_e=-1,\,z_p=1$ and $k_{ij}$ is a cut-off 'momentum' defining a
 diffraction correction allowing for
quantum effects. In the simplest formulation $k_{ij}$ is the thermal de
 Broglie momentum given by:
\begin{eqnarray}
\label{kdifr}
k_{ij}&=&k^{dB}_{ij}=(2\pi\, m_{ij} T_{ij})^{1/2}\\
m_{ij}&=&m_im_j/(m_i+m_j)
\end{eqnarray}
The temperatures $T_{ij}$ entering into the HNC equations for the $g_{ij}(r)$
are given by:
\begin{equation}
\label{crosstemp}
T_{ij}/m_{ij}=T_i/m_i+T_j/m_j.
\end{equation}
 The large mass of the proton ensures that the diffraction
 correction, as well as the $T_q$, is negligible
for the proton-proton scattering process. A more complete approach to
 determining $k_{ij}$ is to
solve the corresponding Kohn-Sham equation for the two particles in the
 Kohn-Sham potential of the
medium, and matching the $k_{ij}$ so that the quantum and classical values
of the PDF at contact are in agreement. 

The Pauli exclusion potential is usually determined by inverting the
 HNC equations applied to the exactly known
non-interacting quantum PDFs $g^0_{ii}(r)$ of the uniform electron fluid.
Here $i$ runs through $e\uparrow, e\downarrow$,
i.e., a spin-resolved, two-component electron system is used. If $i\ne j$, $P_{ij}=0$. 
In the absence of strong magnetic fields,
the spin-resolution is not needed in warm dense systems.
Treating the electrons as a spin-averaged 
one-component subsystem simplifies the
simulations.
Due to the non-linearity of the inversion of the HNC relations given
 in Eqs.~\ref{paupot},
 the Pauli exclusion potential
for paramagnetic electrons is not a simple average of the Pauli exclusion
 potentials of the spin-resolved cases. 
 The corresponding Pauli potential
$P_e(r)$ has to be extracted directly from the averaged $g^0_{ee}(r)$.
 Thus we have, for the spin-resolved and -unresolved cases:
 \begin{eqnarray}
 \label{paupot}
 \beta P_{ii}(r)&=&-\ln[g_{ii}^0(r)]+N_{ii}^0(r), \\
 \beta P_e(r)   &=&-\ln[\{g_{ii}^0(r)+1\}/2]+N^0_{ii}/2 \\
 N_{ii}^0(r) &=& h_{ii}^0(r)-c_{ii}^0(r).
 \end{eqnarray}
A bridge term $B_{ij}(r)$ is used to correct the HNC for multi-particle
 effects poorly rendered by HNC. Such bridge corrections are found to be
 very significant in the 2D electron fluid\cite{prl2},  but not
for 3D electrons at densities and temperatures considered in this study.

The CHNC differs from HNC in the use of the Pauli potentials and the
 quantum temperature $T_q$  when treating quantum fluids. Also, the pair-potentials
 used in HNC have been constructed to agree with KS-charge profiles (see below).
  Hence any insights obtained for the two-temperature two-mass HNC can be
easily transfered to the CHNC. The two-temperature electron-ion
plasma was discussed in a formal analysis by Boercker and More\cite{bm}, using
a  product form for the partition function. However, no
comparisons of their results with actual simulations are available. 
The more general two-temperature two-mass HNC type equations
 have been discussed, most recently by Seuferling et
  al.\cite{seuf}. Using an analysis based on the 
  Bogoliubov-Born-Green-Kirkwood-Yvon (BBGKY) hierarchy as
   well as some factorization assumptions, the authors of 
Ref.~[\onlinecite{seuf}] have proposed modified OZ type
 equations for 2T-2M plasmas. While their formulae reduce to the usual
  OZ equations, viz., Eq.~\ref{oz1}, for the $m_a>>m_b,\; T_a=T_b$, the
   case $m_a=m_b,\; T_a=T_b$ is not correctly recovered. The results
  presented in our study imply that the usual OZ equations hold in all cases,
 as long as the correct mass-dependent $T_{ij}$, Eq.~\ref{crosstemp}
  is used in 2T-2M systems. 
\subsection{The Kohn-Sham reference calculation}
\begin{figure}
 \includegraphics*[width=8.8 cm, height=11.0 cm]{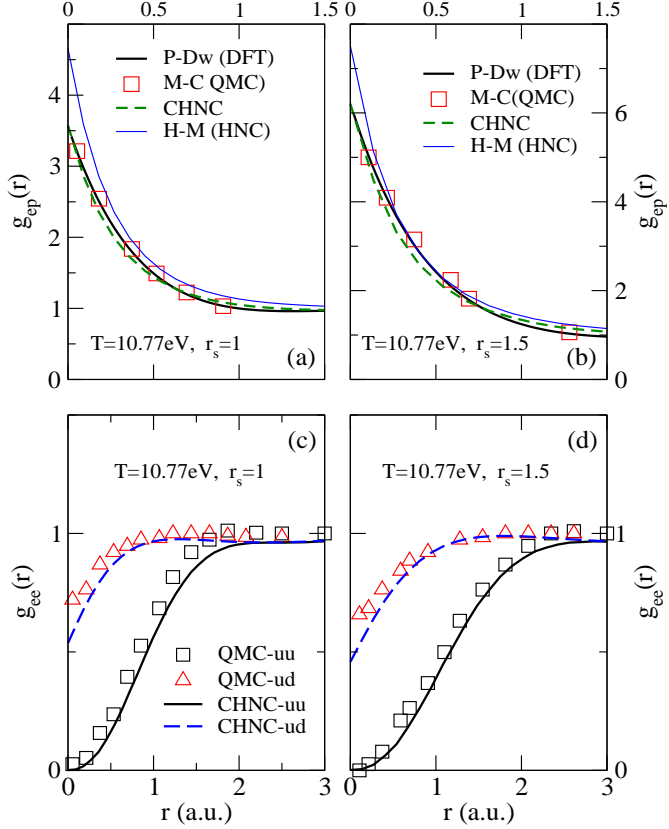}
 \caption
{(Online color)Panels (a),(b) present a comparison of the DFT, QMC,
 CHNC, MD, and HNC calculations of the electron-proton PDF.
  The last three use the effective potentials
 of Eq.~\ref{effpots}. A quantum temperature $T_q$, a Pauli potential
 and the effctive potentials with $f_{ep}$ different from unity are used in
 CHNC. The lower panels compare the spin-resolved 
 electron-electron PDfs in the H-plasma, obtained from QMC and CHNC.
 P-Dw(DFT) and CHNC calculations use the formulations of Dharma-wardana
  and Perrot\cite{ilciacco, prl1}. The
 QMC (PIMC) is from Militzer and Ceperley\cite{milicep},
 while the H-M(HNC) follow the MD calculations of
 Hanson and MacDonald\cite{hanmac} using the potentials of Eq.~\ref{zerothset}.   
 }
\label{ksqmc}
\end{figure}

Kohn-Sham theory at finite temperatures\cite{ilciacco} states that
 the true one-particle density distribution of the system subject to an
  external potential is such that the free
 energy of the system is minimized. This theorem holds rigorously for
  a system in equilibrium and we use it to derive distribution functions
  by considering the inhomogeneous electron distribution
 around a proton in the plasma. Let $n(r)$ and $\rho(r)$ be the electron
  and proton charge densities around the proton at the origin. These tend
   to the average densities $\bar{n}=\bar{\rho}$ far
 away($r\to \infty$) from the proton at the center; then,
 \begin{equation}
 g_{ep}(r)=n(r)/\bar{n}.
 \end{equation}
 Instead of using a two-component DFT procedure, we make the further
  approximation, well established
 in practice, where the proton subsystem is replaced by a uniform positive
  background with a cavity, viz., a Wigner-Seitz sphere of radius
   $r_s=\left[3/(4\pi\bar{n}\right]^{1/3}$. The positive charge scooped out
 to form the  cavity is placed as a point charge at the origin, and forms
the central proton. The finite-temperature Kohn-Sham equation is a
 consequence of the Euler equation for the stationary property of the free
  energy under functional derivation with respect to the
 electron-density distribution, {\it viz}.,
 \begin{equation}
 \frac{\delta F([n(r)])}{\delta n(r)}=0.
 \end{equation}
 A standard Kohn-Sham type analysis now leads to the equation:
 \begin{equation}
 \label{kohnsham}
 [-\nabla^2/2+Z/r-V_{ks}(r)]\psi_\nu(r)=
 \epsilon_\nu\psi_\nu(r),
 \end{equation}
 where
 \begin{eqnarray*}
 V_{ks}(r)&=&V_p(r,n(r))+V_{xc}(r,n(r),T_e)\\
 n(r)&=&\sum_\nu |\phi_\nu(r)|^2f(\epsilon_\nu/T_e).
 \end{eqnarray*}
 Here
  $$V_P(r,n(r))=\int d{\bf r^\prime}n(r^\prime)/|{\bf r}-{\bf r^\prime}|$$
  is the Poisson potential of the electron distribution $n(r)$. This
   distribution is evaluated self-consistently from the Kohn-Sham
    wavefunctions $\psi_\nu(r),\;\nu=n,l,m$,
  energy $\epsilon_\nu$, with the occupation factor
   given by the Fermi function $f(\epsilon/T_e)$.
   The potential
  due to the proton at the origin is $Z/r$, with $Z=1$,
   and $V_{xc}(r,n(r),T_e)$ is the finite-temperature Kohn-Sham
   exchange-correlation potential\cite{prb3d} which depends self-consistently on
   the charge profile $n(r)$. This is evaluated using the local-density
    approximation (LDA), unlike in CHNC where a fully non-local $V_{xc}(r)$
  is evaluated via a coupling-constant integration over the $g_{ee}(r)$.
  The Kohn-Sham procedure uses only $n(r)=\bar{n}g_{ep}(r)$, and does not
  provide a $g_{ee}(r)$. Since this problem contains only one proton, there
  is no proton temperature in the theory. However, due to the
  large mass of the proton, and due to the exclusion of other protons by the
   central proton (modeled by the Wigner-Seitz cavity), the value of
    $g_{ep}(r)$ at $r\to 0$ given by the
  Kohn-Sham calculation is expected to be a valid estimate for the full
   electron-proton plasma. In fact, in the two-temperature electron-proton
    plasma, $T_{ep}$ of Eq.~\ref{crosstemp}
    reduces to $T_{ee}$, as in the Kohn-Sham calculation,
     since $m_p>>m_e$. That this one-proton Kohn-Sham calculation correctly
     reproduces the $g_{ep}(r)$ of the plasma, even at very low temperatures,
     is seen from the comparisons given in Fig.~\ref{ksqmc}, where the path-integral Monte
Carlo (PIMC) PDFs for hydrogen from the work of Militzer and Ceperley\cite{milicep}
     have been used.
\subsection{The effective electron-proton interaction
   for classical simulations.}
  In a classical simulation of an electron-proton plasma, or in a CHNC
   calculation, the Coulomb interaction $V_{ep}(r)$ appears. This is the
    attractive classical potential associated with the quantum-mechanical
     operator $-1/r_{ep}$. It is expected to have the form:
  \begin{eqnarray}
  \label{effpots}
  V_{ep}(r)&=&-\left[1-e^{-k_{ep}r}\right]/r\\
  k_{ep}&=&k^{dB}_{ep}f_{ep}\\
  k^{dB}_{ep}&=&\left[2\pi T_{ep}m_{ep}\right]^{1/2}
  \end{eqnarray}
  The thermal de Broglie momentum $k^{dB}$ 
  provides a first approximation to $k_{ep}$.
   But we choose $k_{ep}$
  such that the $g_{ep}(r\to 0)$ generated by the classical procedure (e.g., MD
   or CHNC) agrees with the  $g_{ep}(r\to 0)$ obtained from the Kohn-Sham
  calculation at the given  $r_s$ and $T_e$.
   It turns out that the correction factor $f_{ep}$ is
  quite close to unity for sufficiently high temperatures.
   Even at $T$=10.77 eV, $r_s=1$,
  i.e., $T/E_F$=0.215, $f_{e}$=0.922 and we see from Fig.~\ref{ksqmc} that the
   agreement between QMC, DFT, and CHNC is quite good. We have 
   determined the value of $f_{ep}$
   as a function of $T_e, r_s$ by matching the CHNC calculation
   and the Kohn-Sham calculation. In effect, $f_{ep}$ is
   similar to a pseuodopotential or form factor for the electron-proton
   interaction. When bound states begin to be formed ($r_s>1.8$), the form of
   $f_{ep}$ becomes more critical, but this problem does not arise within
   the densities studied here. However, classically, for attractive
   potentials, dynamical instabilities could occur at any $r_s$, $T$
   and these have to be controlled using close-approach cutoffs on
	the potentials, as well as controls on the
        velocity distribution functions, to maintain the meaning
        of ``subsystem temperatures'' which are set to $T_e$ and $T_i$.
        Investigation of such instabilities where the velocity
        distributions do not conform to the two-temperature model
        is outside the scope of this study.
 
   The electron-electron interaction used in
   CHNC, and MD simulations,
  is also a diffraction-corrected Coulomb potential,
  $V_{ee}(r)$, with $k_{ee}$ being
  $(2\pi m_{ee} T_{cf})^{1/2}$, with $m_{ee}=m_e/2$ and requiring no
  additional correction factors.
   This
   diffraction correction
  can be derived from the
  Schr\"odinger equation describing electron-electron scattering\cite{prl3}. 
  \subsection{Classical-map Molecular dynamics}
  The HNC method using all three items: (i) diffraction-corrected effective potentials,
   (ii) the  Pauli exclusion potential and (iii) the quantum temperature $T_q$,
   is the CHNC scheme. If the same three items were included in classical
   molecular dynamics simulations, we have a classical-map-MD scheme (CMMD).
   The CMMD is superior to CHNC since it
   automatically includes any bridge corrections, etc. 
 that are not in the HNC scheme. In CMMD the electron temperature would be $T_{cf}$,
 Eq.~\ref{tcf}, as in CHNC. However, since bridge corrections are expected to be
 negligible in the regime of densities and temperatures considered here,
 we do not carry out CMMD simulations.
  \section{Results}
  We first provide comparisons between simple classical MD simulations
  and HNC calculations of PDFs of 2T-2M systems using the simplest diffraction corrected
  pair potentials, given by: 
  \begin{eqnarray}
  \label{zerothset}
  \phi^0_{ij}&=&z_iz_j(1-e^{-k^{dB}_{ij}r})/r\\
  k^{dB}_{ij}&=&\left[2\pi m_{ij}T_{ij}\right]^{1/2}
  \end{eqnarray}
  The MD simulations only need the individual subsystem temperatures $T_{ii}$, $T_{jj}$, and
  no cross-species temperature $T_{ij}, i\ne j$ is needed.  This is achieved by employing
two velocity-scaling thermostats that adjust the electron and ion velocity distributions to 
have the desired mean values. In contrast, the HNC needs a
  specification for $T_{ij}$. Seuferling et al.\cite{seuf} have suggested that
   the OZ equations also
  need to be modified. These issues can be tested by comparison with the MD results.
\subsection{Two-mass two-temperature systems.}
Systems where the two masses $m_a$ and $m_b$ of the two components $a,b$
are equal cannot produce two-temperature
quasi-equilibrium systems unless $V_{ab}$ is, for some reason, extremely different from
$V_{bb}$ and $V_{aa}$. Thus two-temperature plasmas may exist for significant times,
even when the mass-ratio is
of the order of 3-10, as in some solid state electron-hole plasmas where band-structure
effects associated with the existence of indirect gaps introduce restrictions on
electron-hole recombination. Here we present HNC
calculations of the PDFs of plasmas with $m_b/m_a=5$,
 and compare them with MD simulations,
to establish the correct implementation of HNC and CHNC procedures.

\begin{figure}
 \includegraphics*[width=8.5 cm, height=11.0 cm]{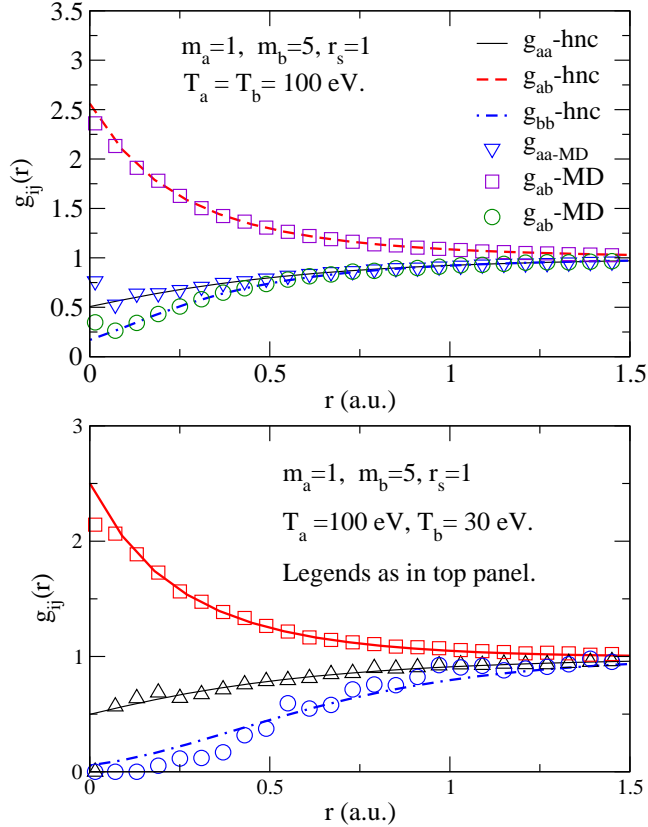}
 \caption
{(Online color)The upper panel shows the PDFs for a two-component ($a,b$) 
system at a temperature $T_a=T_b=100$ eV, $r_s=1$, the masses $m_a,\,m_b$ being 1 and 5.
 The component $a$ electron like, while $b$ is hole-like, i.e., positively charged.
 Lower panel shows the two-temperature two-mass case, with $T_b$ lowered to 30 eV.
 In both cases the PDFs $g_{ij}(r)$ calculated using the standard HNC and the
 OZ equation, Eq.~\ref{oz1}
  agree well with the MD results using the same input potentials as in HNC.}  
 \label{tttm}
\end{figure}

In Fig.~\ref{tttm} we show the PDFs calculated for a two-component system with a mass
ratio of 5, using both the HNC with the standard OZ relations and MD.  The MD simulations used 300 
particles, 40,000 equilibration steps, a time step of 0.02 of the inverse electron-plasma
 frequency, and data was then accumulated over 120,000 steps using the two thermostats described 
above. In the HNC calculation, the  pair-potentials are given by Eq.~\ref{zerothset}, and the 
cross-species temperature is  as in Eq.~\ref{crosstemp}.
This simple HNC-OZ procedure is in very good agreement with
 MD, both for the equilibrium and non-equilibrium (two-temperature) cases, and we
 conclude that the additional procedures proposed by Seuferling et al.\cite{seuf} in
 their Eq. (37) are not needed. That is, our results show that a modified OZ equation is not
necessary.  
 The comparison between the HNC and the MD establishes the correctness
 of the basic HNC procedures even in the quasi-equilibrium case where the formal
 derivation of the HNC equations becomes an open question.
 However, once the correct HNC procedure is established, the
 calculations for the quantum two-temperature two-mass system can be carried out using
 the CHNC, with the same temperature assignments $T_{ij}$ extended to include the
 quantum temperatures $T_q$, and the Pauli potentials.
\subsection{Electron-proton systems in thermal equilibrium}
  The simple diffraction corrected potentials, Eqs.~\ref{zerothset}
   were used by Hanson and MacDonald in H-plasma
  simulations\cite{hanmac}. Their PDFs can also be generated using the
  simple HNC equations if the above $\phi_{ij}(r)$ are used. Hence, in
  Fig.~\ref{ksqmc} we have labeled the corresponding $g_{ep}(r)$ as H-M(HNC,MD).
  The PDFs obtained from the DFT calculation, using Eq.~\ref{kohnsham},
   as implemented in the
  codes by Perrot and Dharma-wardana\cite{dpcodes}, as well as the PIMC
  results of Militzer and Ceperley are also shown, to establish that these
  two first-principles methods are in excellent agreement. Here we note
  that the CHNC results for $g_{ep}$ and also the spin-resolved $g_{ee}$ are in
  excellent agreement with the QMC PDFs. To obtain this agreement,
   the CHNC uses the slightly modified
  diffraction parameter $k_{ep}=f_{ep}k^{dB}_{ep}$ with $f_{ep}$=0.922 and 0.965
   at $r_s$=1 and 1.5 respectively, as obtained by matching the contact value
  of the CHNC $g_{ep}$ to the value from the KS calculation.
  The MD-HNC using the Hanson-MacDonald 
   approach leads to a large value of $g_{ep}$ at $r\to0$, while the $g_{ee}$
   (not shown in the figure) are in strong disagreement.
   The agreement between QMC and CHNC shown in Fig.~\ref{ksqmc} holds even
   better at higher temperatures, and this justifies our use of the CHNC
   and Kohn-Sham results as the reference calculations when QMC
   results are not available.
   \subsection{Two-temperature electron-proton systems}
\begin{figure}
 \includegraphics*[width=9 cm, height=12.0 cm]{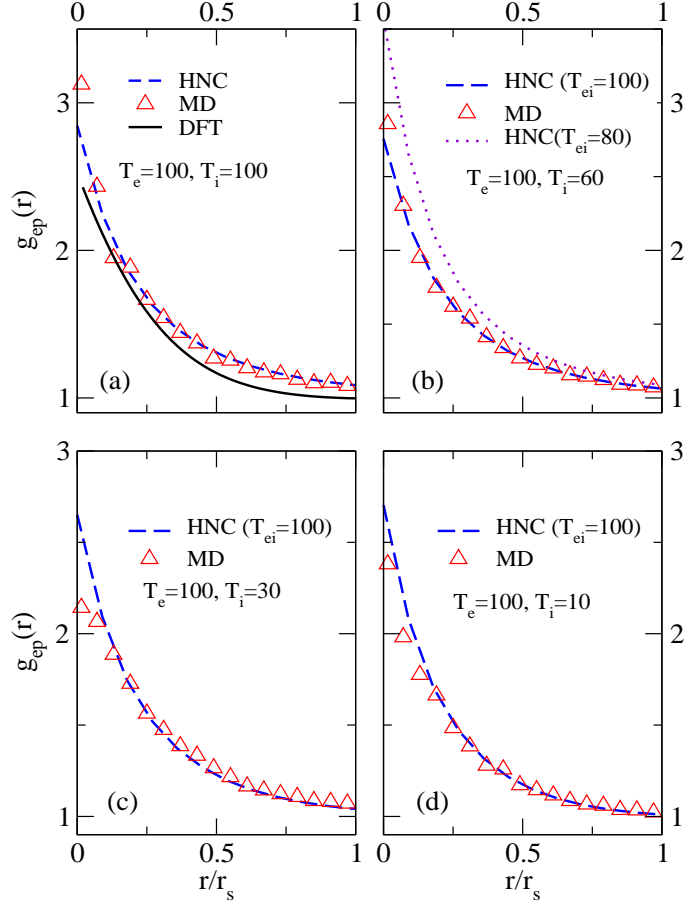}
 \caption
{(Online color) Dense hydrogen: Panel (a) compares HNC and MD $g_{ep}(r)$ using the simplest set
of classical potentials (Eq.~\ref{zerothset}). The DFT PDF shows that the
classical potentials are an overestimate. In panels (b-d) we use the
same classical potentials to establish that the temperature $T_{ep}$ 
needed in the HNC is indeed $T_e$ if HNC and MD are to agree for
two-temperature electron-proton systems.}
\label{mdhnc}
\end{figure}

   In this sub-section we compare classical two-temperature H-plasmas and show that
  the temperature $T_{ep}$ that appears in the cross-species HNC equation is
  indeed the electron temperature $T_{ee}$, as in Eq.~\ref{kdifr},
   for the limit $m_p>>m_e$. Thus, we use the same $T_{ep}$ in the CHNC,
    to include
  the quantum corrections and compute $g_{ij}(r)$.
   In Fig.~\ref{mdhnc} we
  show the cross-species (electron-proton) PDF for a hydrogen plasma with
  $T_e=100$ eV, $r_s=1$, for the four cases:  $T_i=100,\,60,\,30$ and 10.
   From panel (a)
  we see that the classical procedures (HNC and MD) using the simplest
  set of $\phi_{ij}(r)$, Eq.~\ref{zerothset}, overestimate the $g_{ep}$
   in comparison
  to the Kohn-Sham (DFT) estimate. In panels (b-d) we have two-temperature
  plasmas, and the MD calculation (which needs only $T_e,T_i$) is
  closely reproduced by the
  HNC if $T_{ep}$ is set to $T_e$. In panel (b) we show that the choice
   $T_{ep}=(T_e+T_p)/2$ in the HNC is clearly inapplicable if the system
   is entirely specified by $T_e,\,T_p,\,$ and $r_s$.
  As seen from Fig.~\ref{mdhnc}, quantum effects may significantly modify the
  PDFs even when the electrons are at 100 eV. 
  
\begin{figure}
 \includegraphics*[width=9 cm, height=12.0 cm]{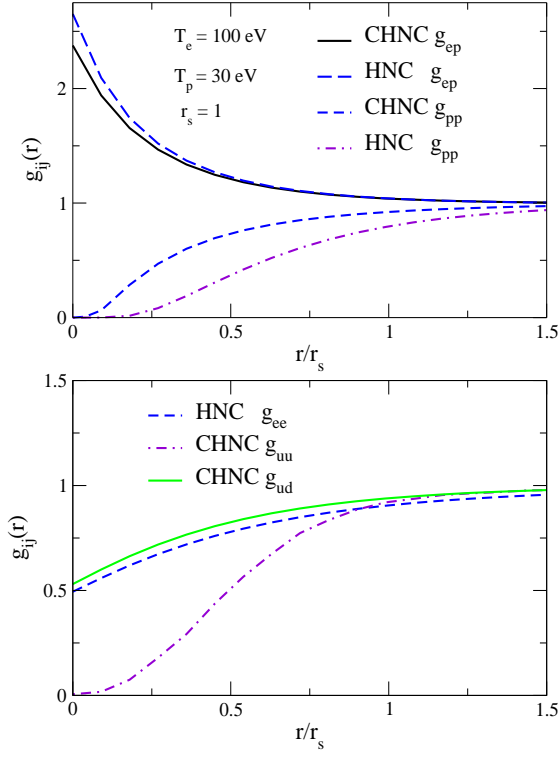}
 \caption
{(Online color)The upper panel shows the e-p and p-p PDFs for
an electron-proton plasma with $T_e=100$eV, $T_p=30$eV, $r_s=1$, calculated using
HNC and CHNC. The lower panel shows the e-e PDFs, where the HNC does not
incorporate the effects of the exclusion principle. The CHNC $g_{uu}$ and
$g_{ud}$ refer to spin parallel and antiparallel PDFs respectively.}  
 \label{chnc2t}
\end{figure}

  Thus, in Fig.~\ref{chnc2t}, we present
  CHNC calculations for a two-temperature plasma with $T_e=100, T_i=30$ at the
  density $r_s=1$. The top panel shows that the proton-proton PDF calculated from
  the quantum procedure (using CHNC) is more strongly coupled than in the classical
  (using HNC) $g_{pp}$. The stronger e-p interaction in the classical system,
  as shown in the enhanced $g_{ep}$, leads to greater screening,
   weakening the ion-ion
  interaction. The lower panel shows the spin-resolved e-e PDFs, labeled $g_{uu},g_{ud}$
  obtained from CHNC, and the classical $g_{ee}$ obtained from HNC. The CHNC correctly
  incorporates the exclusion effects via the Pauli potential, Eq.~\ref{paupot}.
\subsection{The electron-proton PDF and Pauli exclusion effects}
The electron-proton pair distribution function is mainly determined by the e-p interaction
which is spin-independent. However, once an electron is correlated with a proton, the
correlation of that electron with other electrons would be affected by Pauli exclusion 
effects associated with the electron spin. In the CHNC and CMMD schemes, the
effect of the Pauli principle are incorporated as a potential, Eq.~\ref{paupot},
 between parallel-spin electrons. This potential is not used in MD and in
 the pure HNC scheme. Hence the
 $g_{ee}(r)$ obtained from HNC,
  shown in the lower panel of Fig.~\ref{chnc2t}(b) is identical for
 parallel and anti-parallel PDFs. However, Fig.~\ref{chnc2t}(a) shows that the $g_{ep}$
 obtained by the full CHNC, inclusive of the Pauli potential, $f_{ep}$, and $T_q$ features
 is quite close to the pure-HNC result where $f_{ep}=1$ in the
 diffraction potentials. At $r_s=1$, $T_q/E_F=0.768$, and hence, when $T_e=100$ eV, i.e.,
 $T_e/E_F=1.9956$, then $T_q$ itself is substantial. Thus the larger value of $g_{ep}(r)$ at
 $r\to0$ found in the HNC and MD is not due to the Pauli exclusion effects,
 but due mainly to two reasons: (i)the overestimate contained in
 the zeroth set of effective potentials where $f_{ep}=1$, and
  (ii)the use of the physical temperature $T_e$ as the effective temperature
 of the classical electron fluid,
  while $T_{cf}>T_e$ is used in the CHNC.
   To check these, we have run CHNC calculations  where (i) the Pauli
 potential was switched off while the $T_q$, $f_{ep}$ were included; (ii) only the Pauli 
 and $f_{ep}$ were included; (iii) only the Pauli and $T_q$ were included; and so forth.
 Such ``numerical experiments'' enable us to conclude that the Pauli exclusion effect is
 of relatively low importance for the $g_{ep}(r)$ when $T_e$ is 100 eV and $r_s=1$. 

\section{concluding discussion}
 The simplest classical rendering of
 quantum plasmas, based on the use of diffraction corrected potentials
 (Eq.~\ref{zerothset}) was used with HNC calculations and MD simulations to resolve 
 the ambiguities and difficulties in handling the two-temperature, two-mass system.
 We conclude that the modifications to the OZ equations proposed by
 Seuferling et al.\cite{seuf}., are not needed.
  The classical
 mapping of quantum systems to the HNC equations, as used in the CHNC was confirmed by
 comparisons with Kohn-Sham DFT calculations as well as with available PIMC
 results for compressed hydrogen plasmas at finite temperatures.
 We conclude that the HNC and CHNC, together with the standard OZ
 equations provide
  excellent, accurate and simple analytical tools for the investigation
 of many-particle quasi-equilibrium
  systems for which direct quantum simulations continue to remain too prohibitive
 or unfeasible.   


\begin{thebibliography}{99}
%
\bibitem{ceperleyH}
 K. T. Delaney, C. Pierleoni, D. M. Ceperley. Phys. Rev. Lett. {\bf 97}, 235702 (2006). 
 C. Pierleoni, D. M. Ceperley, Markus Holzmann. Phys. Rev. Lett. {\bf 95}, 146402 (2004)
\bibitem{kwon}
I. Kwon, L. Collins, J. Kress and N. Troullier, Phys. Rev. E. {\bf 54},
2844 (1996)
 \bibitem{dejarlais}
M. P. Desjarlais, Phys. Rev. B {\bf 68}, 64204 (2003);
\bibitem{mazevet}
S. Mazevet, M. P. Desjarlais, L. A. Collins, J. D. Kress and N. H. Magee,
Phys. Rev. E {\bf 71}, 016409 (2005)
\bibitem{pdw95}
F. Perrot and M.W.C. Dharma-wardana, Phys. Rev. E.  {\bf 52 }, 5352 (1995)
\bibitem{ilciacco}
{\it Density Functional Theory}, Ed. E. K. U. Gross and Dreizler,
NATO ASI Series B: Physics 337 (Plenum, NY. 1993)
\bibitem{lvm}
M.W.C. Dharma-wardana, Phys. Rev. E. {\bf 73},  036401 (2006)
\bibitem{astrop}
F. A. Agronyan and R. A. Syunyaev,
 Astrophysics {\bf 27}, 413-422; Translated from Astrofizika; 27: No. 1, 131-145 ( 1987)
\bibitem{dewitt}
A.I. Chugunov, H. E. DeWitt, and D. G. Yakovlev, Phys. Rev. D {\bf 76}, 025028 (2007)
\bibitem{seuf}
P.Seuferling, J. Vogel, and C. Toepffer, Phys. Rev. A {\bf 40}, 323 (1989)
\bibitem{mc}
D. P. Landau and K. Binder {\it A guide to Monte Carlo simulations in statistical Physics}
(Cambridge University Press 2005)
\bibitem{hncref}
J. M. J. van Leeuwen, J. Gr\"oneveld, J. de Boer, Physica {\bf 25}, 792 (1959)
\bibitem{prl1}
M. W. C. Dharma-wardana and F. Perrot, Phys. Rev. Lett. {\bf 84}, 959 (2000)
\bibitem{prb3d}
F. Perrot and M.W.C. Dharma-wardana, Phys. Rev. B, {\bf 62}, 16536 (2000);
\bibitem{Jones} C. S. Jones and M. S. Murillo,
 High Energy Density Phys. (doi:10.1016/j.hedp.2007.02.038) 
(2007).
\bibitem{feenberg}
E.Feenberg, {\em Theory of Quantum Fluids} (Academic, New York 1969)
\bibitem{lantto}
L. J. Lantto,  Phys. Rev. B {\bf 22}, 1380 (1980)
\bibitem{prl2}
F. Perrot and M. W. C. Dharma-wardana, Phys. Rev. Lett. {\bf 87}, 206404
   (2001)
\bibitem{lado}
 F. Lado, J. Chem. Phys. {\bf 47}, 5369 (1967)
\bibitem{bm}
D. B. Boercker and R. M. More, Phys. Rev A {\bf 33} 1859 (1986)
\bibitem{milicep}
B. Militzer and D. Ceperley , Phys. Rev. Lett. {\bf 85}, 1890 (2000);
B. Militzer, Thesis, (2000), see http://militzer.gl.ciw.edu/diss/diss\_militzer.pdf
\bibitem{dpcodes}
M. W. C. Dharma-wardana and Fran\c{c}ois Perrot;\\
http://athens.phy.nrc.ca/ims/qp/codes/chandre/D\_P/\\
access by password obtainable from: chandre.dharma-wardana@nrc-cnrc-gc.ca
%
\bibitem{hanmac}
J.-P. Hansen, I. R. MacDonald, Phys. Rev. Lett. {\bf 41} 1379 (1978)
\bibitem{prl3}
M.W.C. Dharma-wardana and F. Perrot, Phys. Rev. Lett. {\bf 90}, 136601 (2003)    
  
 
\end{thebibliography}
\end{document}